# The electrical current effect in phase separated $La_{5/8-y}Pr_yCa_{3/8}MnO_3$: Charge order melting vs. Joule heating

Joaquin Sacanell[*], A. G. Leyva and P. Levy

Unidad de Actividad Física, Centro Atómico Constituyentes, CNEA, Av. Gral. Paz 1499

(1650), San Martín, Prov. de Buenos Aires, Argentina.

We have studied the effect of electric field on transport properties of the prototypical phase separated manganite $La_{5/8-y}Pr_yCa_{3/8}MnO_3$ with y=0.34. Our results show that the suggested image in which the charge ordered state is melted by the appliance of an electric current and/or voltage has to be revised. We were able to explain the observed resistivity drop in terms of an artifact related to Joule heating and the particular hysteresis that the system under study display, common to many other phase separated manganites.

75.47.Lx, 64.75.+g, 71.30.+h

---

[*] Corresponding author: sacanell@cnea.gov.ar

I. Introduction

Manganese binary oxides, also known as manganites have been the focus of extensive research since the discovery of the colossal magnetoresistance (CMR) effect[1], i.e. the reduction of the resistivity by several orders of magnitude in the presence of an external magnetic field. In the recent literature, an important amount of work was devoted to study the particular low temperature state of some manganites characterized by the intrinsic coexistence of two or more phases on a submicrometric scale [2], a phenomenon named phase separation (PS). As thoroughly documented [2,3,4], this mixture is usually formed by a ferromagnetic (FM) conductive and a highly insulating charge ordered (CO) phases. It is now believed that the CMR effect is closely related to the increase of the FM fraction induced by the magnetic field[5].

Many efforts have been performed in order to understand the effect of other stimuli on the PS state, trying to destabilize one or the other phase by the application of hydrostatic pressure[6], thermal cycles[7], X ray[8] or laser[9] irradiation, and electric field [10,11,12,13,14,15,16]. In the last case, the observed effects seem to indicate that the CO state melts under the application of an electric field. However, there is an inherent experimental problem that can hide the true nature of the results, Joule heating will always be present producing undesired thermal changes to the sample. The role of Joule heating is briefly mentioned by several of the authors, but no deep analysis has been made on whether the observed results correspond to electrically induced charge delocalization or to thermal artifacts.

In the present work we study the effects of electric fields in the PS manganite $La_{5/8-y}Pr_yCa_{3/8}MnO_3$ with y=0.34, paying special attention to discriminate whether they are related to the melting of the CO or to heating of the sample. The compound under study belongs to the well known $La_{5/8-y}Pr_yCa_{3/8}MnO_3$ (LPCMO(y)) family which was deeply studied since the pioneering work of M. Uehara et al.[3]. The system is a prototypical PS manganite which has an extensively documented tendency to form inhomogeneous structures characterized by the coexistence of FM and CO phases in the low temperatures region [3,17,18,19,20].

## II. Experimental

We have performed 4 probe resistivity ($\rho$) measurements on a high quality bulk polycrystalline sample of LPCMO(0.34) synthesized by the Sol – Gel technique. The powder was pressed into a bar with the dimension of $5 \times 1 \times 1$ mm$^3$. The average grain size was estimated to be of the order of 2 µm through SEM microscopy.

## III. Results and Discussion

In figure 1(a) we show $\rho$ vs. *T* data for a LPCMO(0.34) sample measured with 5 different currents (0.01, 0.05, 0.1, 0.5 and 1 mA). All measurements were performed in the same cooling run, changing cyclically among the different current values starting from *i* = 0. The inset of fig. 1(a) shows the temperature dependence of $\rho$ for LPCMO(0.34) measured with $i$ = 0.01 mA displaying strong thermal hysteresis, a signature of the percolative M – I transition. On cooling, we see a steep increase of $\rho$ at $T_{CO} \approx 200$ K, and then the M – I transitions at $T_{MI} \approx 80$ K (120 K on warming).

We see that the resistivity curves for the lower currents ($i$ = 0.01 – 0.1 mA) are almost identical, but $\rho$ is reduced for higher currents. At first sight this fact could be interpreted as arising from the melting of the CO state. The reduction of the resistivity is only observed in the 120 to 70 K range, similar to what is reported by Ma et al. for LPCMO(0.35), which has almost the same composition of our samples[16]. The temperature window in which the effect is happening is characterized by an anomalous hysteresis and unique dynamical effects which have been already reported in other similar manganite systems[21]. These effects are evidenced by an unexpected change of the $\rho$ vs. $T$ slope. In the following, we will show that it is not possible to reject the effect of the Joule heating on the observed phenomenon from these measurements. Moreover, we will show that the reduction of the resistivity is likely to be an artifact related to the anomalous hysteresis described above.

In the measurement of figure 1 (a), due to Joule heating both at contacts and bulk, the sample is forced to perform undesirable minor thermal cycles, as small cooling and warming sweeps are induced by the application and removal of the electric current. This artifact also happens in a typical measurement in which a single electrical current value is turned on and off at each measuring temperature, to avoid an excessive Joule heating.

We have performed an experiment to see whether CO melting is significant or not in our case. In figure 1(b) we show two $\rho$ vs. $T$ curves performed in independent runs, with currents $i = 0.01$ mA and $i = 1$ mA. To avoid the forced cooling and warming cycles at each temperature, we have performed this experiment without turning off the current between measurement points. By following this procedure we assure that the sample's temperature sweep is monotonous.

In the results of figure 1(b), we expect to have large heating effects but, if CO is melting as figure 1(a) seems to indicate, the peak of the resistivity measured with $i = 1$ mA should be smaller than the one corresponding to $i = 0.01$ mA. We see that this is not the case, the two peaks in figure 1(b) reach almost the same value, the only noted difference is the temperature scale. The difference in the peak temperature arises undoubtedly because the measurement performed with the larger current displays greater heating effects than the one measured with the smaller value. But, interestingly, we see that in this case the peak resistivity value has no dependence with the value of the applied current, which means that there is no additional electric field effect other than heating.

The last result is showing that, using the same currents as the ones of the experiment of figure 1(a), no significant melting of the CO is produced. But then, we still need an explanation for the reduction of the resistivity observed in fig. 1(a), which is similar to already reported data, and was ascribed according to the authors, to charge delocalization [15,16].

We are going to show now that the particular hysteresis of LPCMO(0.34)[21], together with the Joule heating effects, can alone account for the reduction of the resistivity observed in the 70 to 120 K range for the compound.

In figure 2 we show a $\rho$ vs. $T$ curve measured following the sequence Room $T \rightarrow$ 80 K $\rightarrow$ 100 K $\rightarrow$ 20 K at a 2 K/min rate. We see that the sample displays a thermal irreversibility related to hysteresis. Surprisingly, $\partial \rho / \partial T$ which is positive while cooling the sample below the M-I transition, becomes negative when the temperature sweep is inverted at 80 K. Measurements of figure 2 were performed with $i = 0.01$ mA, the lowest current used in the present work. We have shown in fig. 1(a) that Joule effects are negligible using $i < 0.1$ mA so we can neglect heating in this experiment.

We see that an inversion of the temperature sweep, if forced below the temperature of the peak ($T_P$), instead of resulting in an increase of $\rho$ (which is expectable following a reversible path) provokes an unexpected reduction of the resistivity. This behavior has been attributed in previous works to dynamical features of the PS state [21]. In particular, it was observed that $\partial \rho / \partial T$ changes its sign if the temperature sweep is inverted below the insulator to metal transition as shown in fig. 2 for LPCMO(0.34).

Through figure 2 we can infer the situation in a typical measurement process: if the sample is at $T = 80$ K and we turn on the current for a short time, local heating is expected to arise from Joule effect. This heating will lead to a reduction of the resistivity as the one shown in fig. 2, which would be larger for larger currents. If the sample is rapidly heated by the application of the electrical current, the thermometer could not be able to detect a fast change in temperature (due to the local nature of heating, velocity of the process, thermal inertia, etc). Then, the reduced resistivity value would be the one registered for $T = 80$ K. So for $T < T_p$ the resistivity curve would be lower if we increase the measurement current. For $T > T_p$ the situation is simpler, as the resistivity curve is reversible in this range, a thermal heating of the sample would obviously lead to a reduction of the resistivity.

The thermal increase can be calculated within a simple image in which a certain power is dissipated in the electrical contact due to its resistance, and heat is transferred to the sample holder through the sample's volume. Then the temperature difference between the top and bottom of the sample is:

$$\Delta T = \frac{P \cdot l}{\kappa \cdot s},$$

where $P$ is the dissipated power, $l$ is the "height" of the sample through which heat is transferred, $\kappa$ is the thermal conductivity[22] and $s$ is the section of the electrical contact. With the values of our experiment and assuming that an equivalent amount of heat is conducted to the sample holder by the measurement wires (which overestimates its value) we have obtained that $\Delta T \approx 25$ K. The temperature window in which the system shows an irreversible behavior is around 30 K (the irreversible zone below $T_{MI}$), so $\Delta T$ is highly significant and enough to screen other effects on the measured values.

Also, a crude estimation of the voltage needed to melt the charge ordered state can be made by comparing the "electrical" energy with the magnetic energy that has to be given to the system for a metamagnetic transition to occur. Typical LPCMO(y) (0.3 < y < 0.4) samples present metamagnetic transitions at low temperature in which the system becomes fully FM by the application of a magnetic field of the order of 2 Tesla[23]. If we suppose that this magnetic field is applied to a 1 $\mu_B$ magnetic moment (which underestimates the 3 or 4 $\mu_B$ of Mn ions) this corresponds to an energy $\varepsilon \approx 2 \times 10^{-5}$ eV per Mn ion. This energy can be compared with the energy $e\,V_l$ needed for an electron to pass from a $Mn^{3+}$ to an adjacent $Mn^{4+}$ to obtain the voltage to be applied to a $Mn^{3+} - O - Mn^{4+}$ union, $V_l$. Then, an electric field around 400 V/cm is needed to melt the CO state. For a typical sized bulk sample ($\approx 1$ cm) the corresponding voltage is 400 V. At 100 K, the resistance of the sample in series with the contact is around 1 k$\Omega$, then a current of 400 mA has to be applied, which is higher than the currents that we have shown to provoke significant heating.

Of course, it would be incorrect to attribute the change of the resistivity in manganites entirely to Joule heating. However, electric fields required to induce the melt of the CO state seems to be in some cases out of the ''safe'' range, and thus special care should be taken while distinguishing it from heating effects.

Joule heating is present in every resistivity measurement and it commonly affects the observed results; however, it is not usually properly taken into account. Even though our results are not enough to discard that electric fields can melt or weaken the CO state in manganites, a doubt could be raised on whether the results are a consequence of charge delocalization or a thermal artifact. In fact, in recent works, several authors have suggested that changes observed in the resistivity of some PS manganites can be attributed to local transitions induced by Joule heating at metallic regions of the sample rather than to electrical effects[24 25 26]. Also, in a very recent work, nonlinearities in *I-V* curves of phase separated $Pr_{0.8}Ca_{0.2}MnO_3$ have been shown to be related with changes in the structural parameters, a result consistent with local heating of the sample around 20 K while applying an electrical current of 5 mA, in excellent agreement with our estimation[27].

**IV. Conclusions**

Summarizing, we have shown that the effect of electrical current on phase separated manganites that present anomalous hysteresis can be explained just by taking into account a thermal increase due to Joule heating which, in turn, produces an unexpected decrease of the resistivity within a definite temperature range. By comparing the two measurement protocols depicted in fig. 1 (a) and (b), we showed that the reduction of the resistivity is an artifact related to the particular thermal irreversibility of the systems under study and not to the melt of the CO state. We have shown that significant heating can occur in typical transport measurements, so these experiments have to be performed taking into account that undesired minor thermal cycles can occur when this current is turned on and off.

**Acknowledgments**

The authors thank M. Quintero, L. Granja, M. Monteverde and S-W. Cheong for fruitful discussions. A. G. Leyva is also at UNSAM, P. Levy is member of CIC CONICET. Support from ANPCyT PICT 03-13517 is acknowledged.

**Figure 1:** (a) $\rho$ vs $T$ cooling the LPCMO(0.34) sample. Measurement was performed cycling from 0.01 to 1 mA while cooling. Inset: $\rho$ vs $T$ for LPCMO(0.34). (b) $\rho$ vs $T$ for LPCMO(0.34) using 0.01 mA y 1 mA, continuously applied during cooling.

**Figure 2:** $\rho$ vs $T$ for LPCMO(0.34) following the temperature sweep indicated by arrows, using $i = 0.01$ mA.

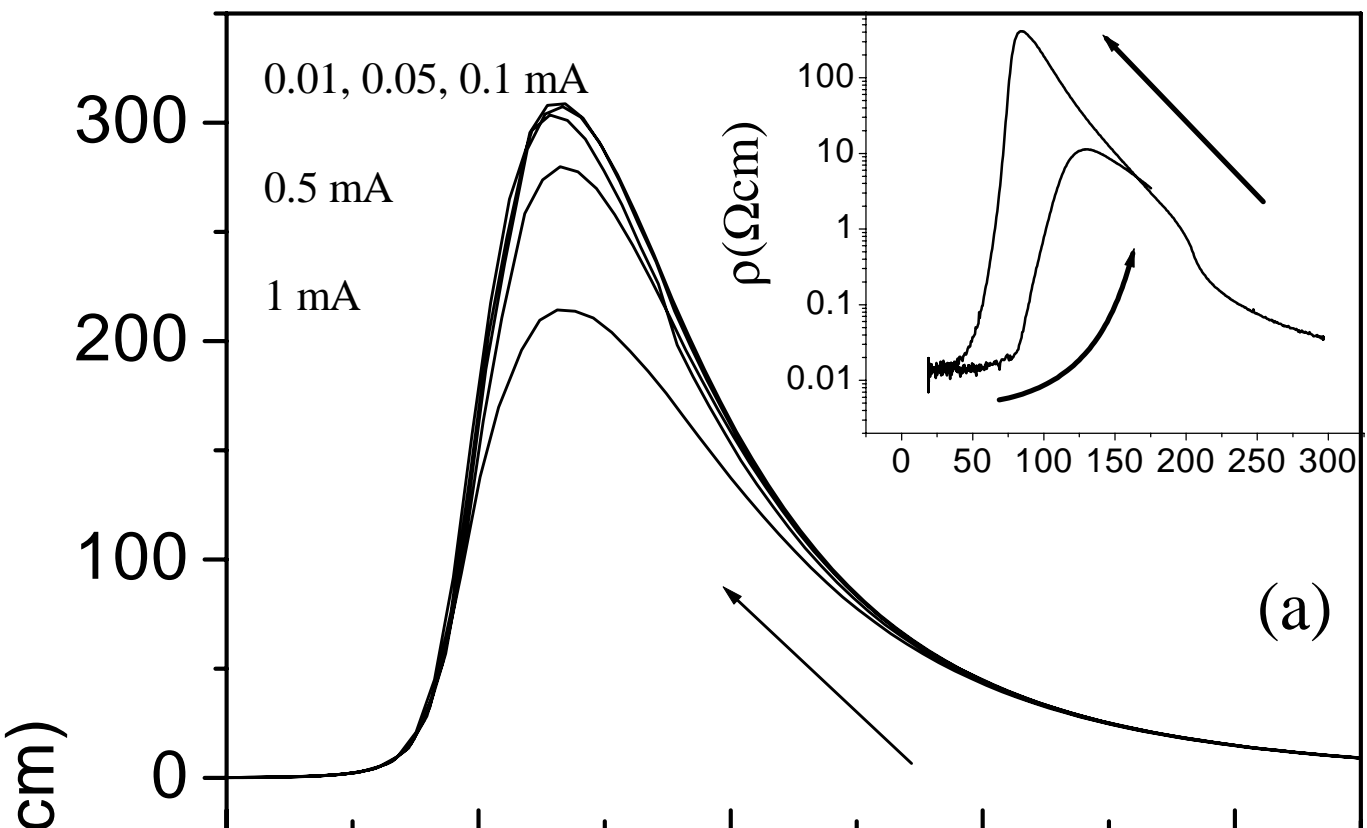

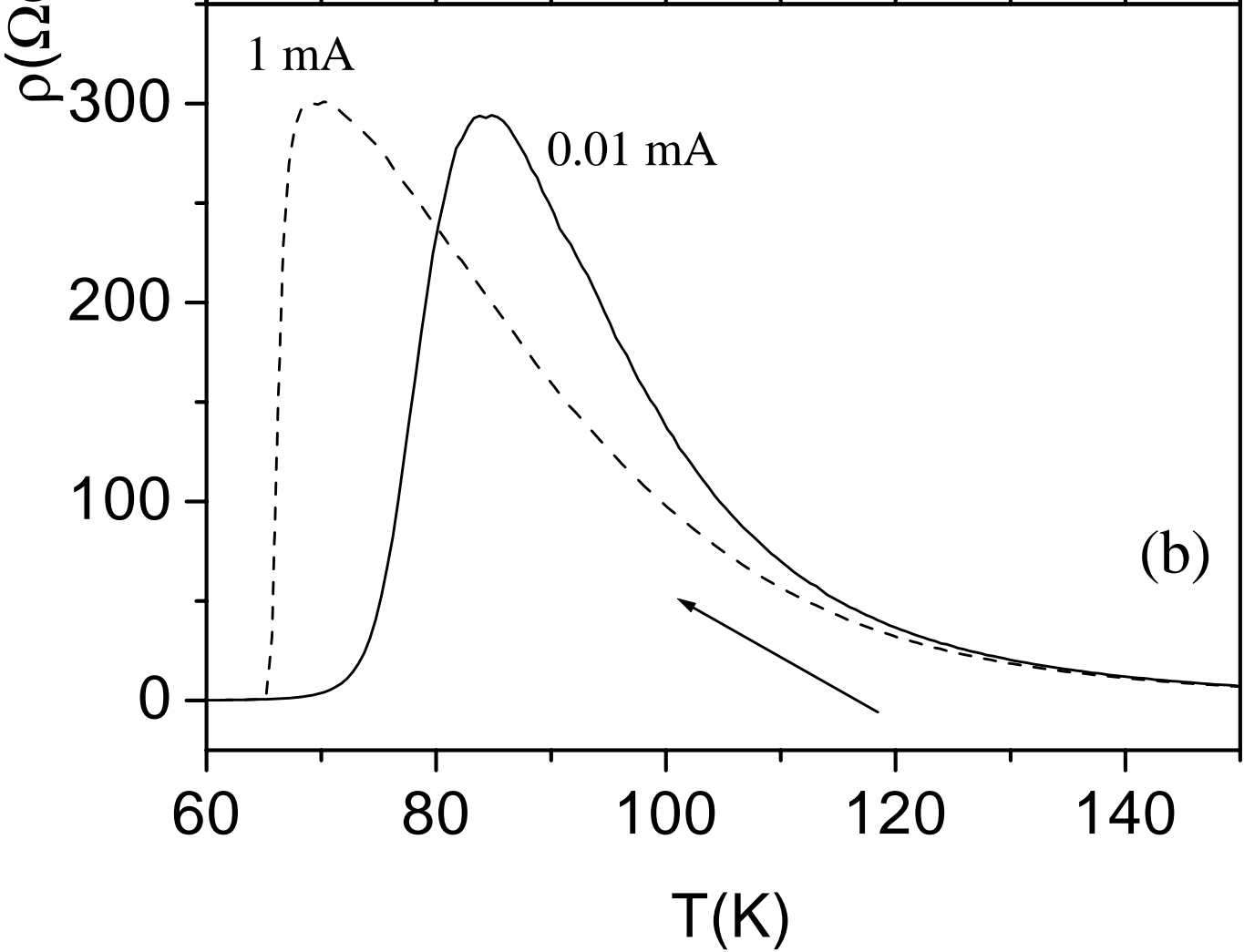

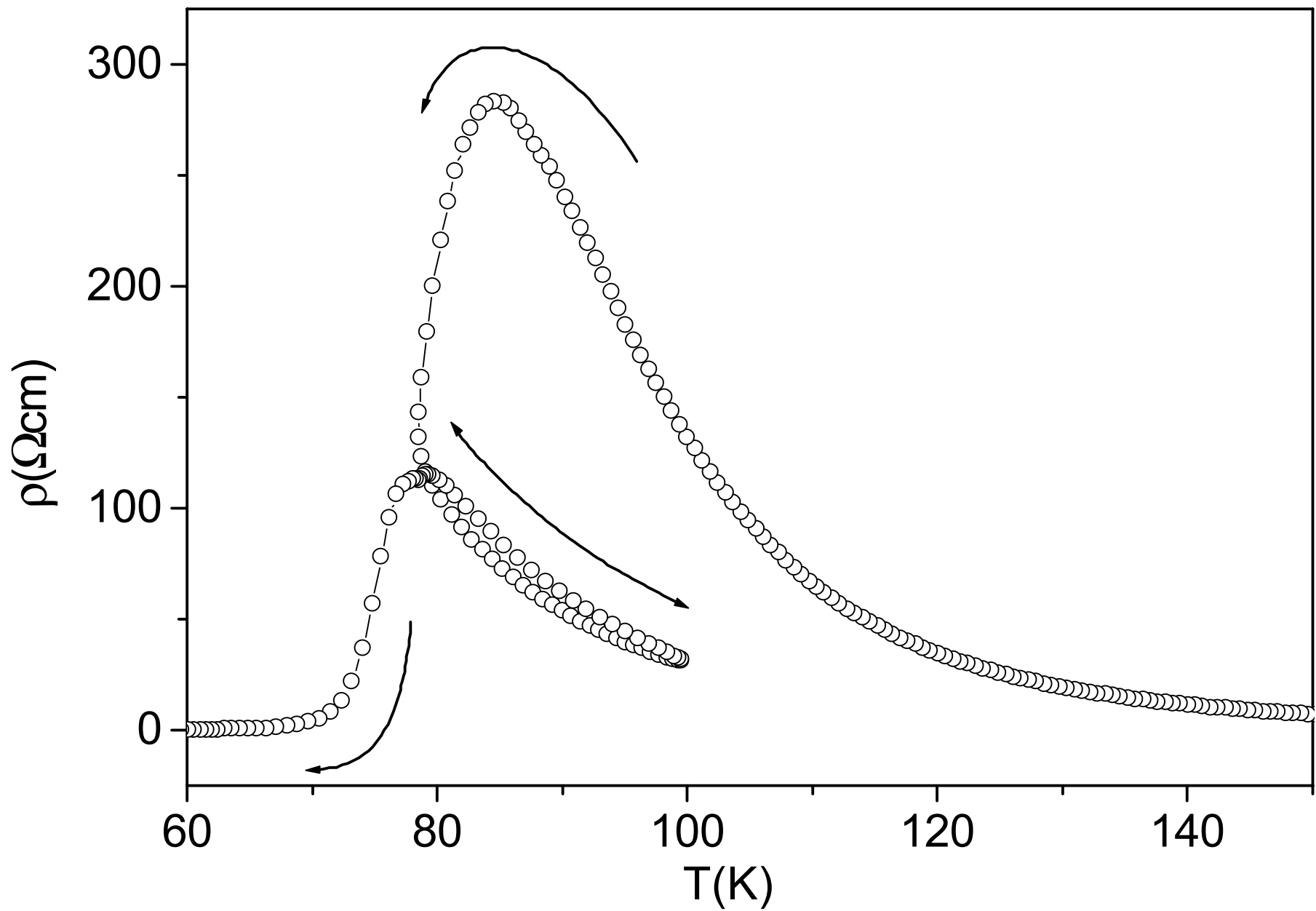